\documentclass[conference]{IEEEtran}
\IEEEoverridecommandlockouts
\usepackage{amsmath,amssymb,amsfonts}
\usepackage{accents}
\usepackage{algorithmic}
\usepackage{graphicx}
\usepackage{textcomp}
\usepackage{xcolor}
\usepackage{xurl}
\usepackage{makecell}
\usepackage{tcolorbox}
\usepackage{floatrow}
\usepackage{balance,cite}
\newfloatcommand{capbtabbox}{table}[][\FBwidth]
\title{Deceptive Jamming in WLAN Sensing
\thanks{This work was supported, in part, by the European Commission through the Horizon Europe/JU SNS project Hexa-X-II (Grant Agreement no. 101095759) and by the Swedish Research Council (VR grant 2022-03007). 
}}
\author{\IEEEauthorblockN{Hasan Can Yildirim\IEEEauthorrefmark{1},
Musa Furkan Keskin\IEEEauthorrefmark{2}, 
Henk Wymeersch\IEEEauthorrefmark{2}, and
Fran{\c{c}}ois Horlin\IEEEauthorrefmark{1}}
\IEEEauthorblockA{\IEEEauthorrefmark{1}Wireless Communications Group, Universit{\'e} Libre de Bruxelles, Belgium}
\IEEEauthorblockA{\IEEEauthorrefmark{2}Department of Electrical Engineering, Chalmers University of Technology, Sweden}}
\begin{document}
\bstctlcite{IEEEexample:BSTcontrol}
\maketitle

\begin{abstract}
Joint Communication and Sensing (JCAS) is 
taking its first shape in WLAN sensing under IEEE 802.11bf, where standardized WLAN signals and protocols are exploited to enable radar-like sensing. However, an overlooked problem in JCAS, and specifically in WLAN Sensing, is the sensitivity of the system to a deceptive jammer, which introduces phantom targets to mislead the victim radar receiver. Standardized waveforms and sensing parameters make the system vulnerable to physical layer attacks. Moreover,  orthogonal frequency-division multiplexing (OFDM) makes deceptive jamming even easier as it allows digitally generated artificial range/Doppler maps. This paper studies deceptive jamming in JCAS, with a special focus on WLAN Sensing. The provided mathematical models give insights into how to design jamming signals and their impact on the sensing system. Numerical analyses illustrate various distortions caused by deceptive jamming, while the experimental results validate the need for meticulous JCAS design to protect the system against physical layer attacks in the form of deceptive jamming.
\end{abstract}


\section{Introduction}
In recent years, joint communication and sensing (JCAS) received attraction from both industry and academia, especially as it constitutes an enabler for 6G \cite{wei2022toward}. JCAS aims at combining communication and sensing capabilities in a single device \cite{JCAS}. While 6G is still several years away, lessons can be learned from one of the first communication-centric JCAS systems under the IEEE Wi-Fi 802.11 standards, namely WLAN Sensing 802.11bf \cite{wlan_sensing}. WLAN Sensing aims to enable presence/intruder and fall detection, identity/gesture recognition, tracking of people, and many more \cite{wlan_sensing_scenarios}. To do so, orthogonal frequency-division multiplexing (OFDM) modulated WLAN signals (at 2.45, 5, 6, and 60 GHz) and the already existing communication-oriented protocols are exploited for radar-like sensing. In particular, standardized training fields, present in the preamble of any Wi-Fi frame to enable channel estimation and equalization for communication, are used in WLAN Sensing \cite{ltf_based_radar}. Therefore, the access points (APs) and user stations (STAs) have access to the original transmit signal, which enables radar-like processing in both monostatic and bistatic geometries \cite{wlan_sensing2}.

\begin{figure}[h!]
    \centering
    \includegraphics[width=0.9\linewidth]{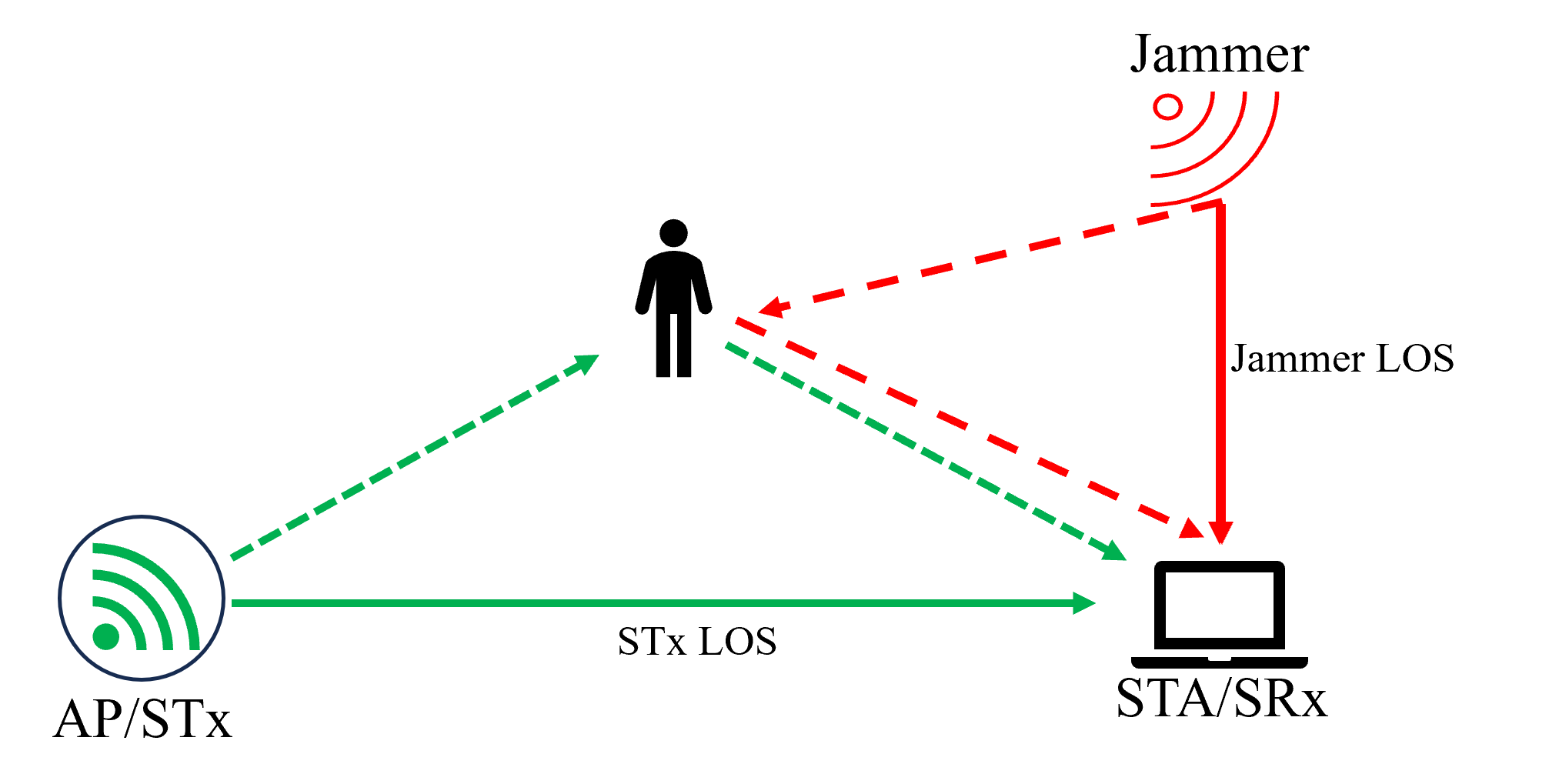}
    \caption{A scenario with a sensing transmitter (STx), sensing receiver (SRx), a mobile target, and a Jammer that transmits pre-modulated signals that carry phantom targets.}
    \label{fig:scenario_geometry}
\end{figure}

Although OFDM has many benefits for JCAS \cite{JCAS_ofdm}, the general JCAS and the WLAN Sensing communities overlook one important fact about using a standardized OFDM waveform for sensing: its sensitivity to deceptive jamming, which aims at misleading a victim radar receiver by generating phantom targets, illustrated in Fig.~\ref{fig:scenario_geometry}. The idea of deceptive jamming is not new in the conventional radar literature \cite{jamming_book}. 
Historically, the deceptive jammer device must (a) estimate the radar operation parameters such as system bandwidth, pulse repetition interval (PRI), etc., (b) reconstruct the original radar waveform with great precision, (c) artificially introduce realistic propagation delays, Doppler shifts, and attenuation to mimic real targets, and (d) transmit these signals within the time frame of the victim radar \cite{jamming_book}. In order to accomplish these in real-time, digital radio-frequency memory (DRFM) technology is often employed where high-speed sampling and digital memory are used to store radar signals \cite{drfm_main}. In the context of wireless sensor networks,  \cite{jamming_survey}  provides an extensive survey about various types of jamming attacks and countermeasures, but ignores the implications related to the use of OFDM waveforms. OFDM deceptive jamming was treated in \cite{ofdm_drfm1,ofdm_drfm2,ofdm_drfm5}. 
In \cite{ofdm_drfm1}, a deceptive jamming framework for OFDM-based synthetic aperture radars (SARs) is proposed, where deceptive target images are injected into the SAR image of the victim radar. Various techniques for reconstructing the original waveform are also discussed. In \cite{ofdm_drfm2},  a new advanced model for airborne deceptive jamming is proposed using DRFM and applying the sub-Nyquist sampling theorem to reduce the sampling rate of the deceptive jammer. Finally, counter-acting deceptive jamming is considered in \cite{ofdm_drfm5}, where the robustness of randomly generated OFDM radar waveforms against deceptive jamming is analyzed. 

JCAS, and more specifically WLAN Sensing, is particularly sensitive to deceptive jamming, since i)
 \emph{radar parameters are selected from a pre-determined set of parameters} \cite{meneghello2023toward}, and ii) the \emph{OFDM waveform is standardized} \cite{wlan_sensing}. These properties remove the need for previously mentioned steps (a) and (b), and hence, the need for DRFM. 
Instead, the primary focus shifts to generating realistic range/Doppler maps (RDMs) and aligning the transmission time within the time frame of the victim radar receiver. The alignment of transmission time benefits from OFDM-based bistatic radar processing which requires the correlation between the transmitted and the received signals to find the timing reference \cite{ofdm_bistatic}. Hence, the timing reference can be triggered by the jamming signal instead of the actual reference signal as long as i) the jamming signal arrives within half of the OFDM symbol duration \cite{nextgen_wlan} and ii) it exhibits a larger amplitude than the actual timing reference signal.
Since JCAS is relatively new in the literature,  deceptive jamming is not yet fully addressed. 
Moreover, the potential and challenges of a deceptive jammer in the JCAS context are not yet validated within a complete system such as WLAN Sensing.

In this paper, we provide insights into what makes WLAN Sensing, or more generally, OFDM-based JCAS, prone to deceptive jamming and analyze various possible scenarios. Our contributions are summarized as follows 
\begin{itemize}
    \item The shortcomings of WLAN Sensing against deceptive jamming are identified, and a mathematical framework to design deceptive jammers is provided.
    \item Thanks to the flexibility of the jamming framework, it is shown that the victim radar can be deceived at the output of the target detector (i.e., per radar snapshot) as well as in the target tracking layer (i.e., over multiple radar snapshots). 
    \item Different deceptive jamming scenarios are discussed, and their impact on the victim radar performance is numerically analyzed.
    \item The vulnerability of WLAN Sensing against deceptive jamming in real-life application environments is experimentally validated using two USRP X310s to emulate the transmitter, the receiver, and the jammer in an indoor scenario with a walking human.
\end{itemize}

\section{WLAN Sensing and System Model}
\label{sec:wlan_sensing}
In this section, we introduce the WLAN Sensing and its characteristics. Then, we provide the  system model (including the OFDM waveform, the channel, and the receiver-side processing).

\subsection{Fundamentals of WLAN Sensing}
In WLAN Sensing \cite{wlan_sensing}, a given Wi-Fi device can be i) a sensing transmitter (STx) which only transmits; ii) a sensing receiver (SRx) which only receives, and iii) a sensing transceiver (STRx) which transmits and receives sensing signals \cite{wlan_sensing2}. These roles simply determine the radar geometries: i) bistatic if STx and SRx are separated devices as illustrated in Fig.~\ref{fig:scenario_geometry}, or ii) monostatic if a single device is acting as an STRx. In this paper, we specifically focus on the bistatic WLAN Sensing geometry since it is the most vulnerable to jamming --with AP and STA acting as STx and SRx, respectively.

\begin{figure}
    \centering
    \includegraphics[width=\linewidth]{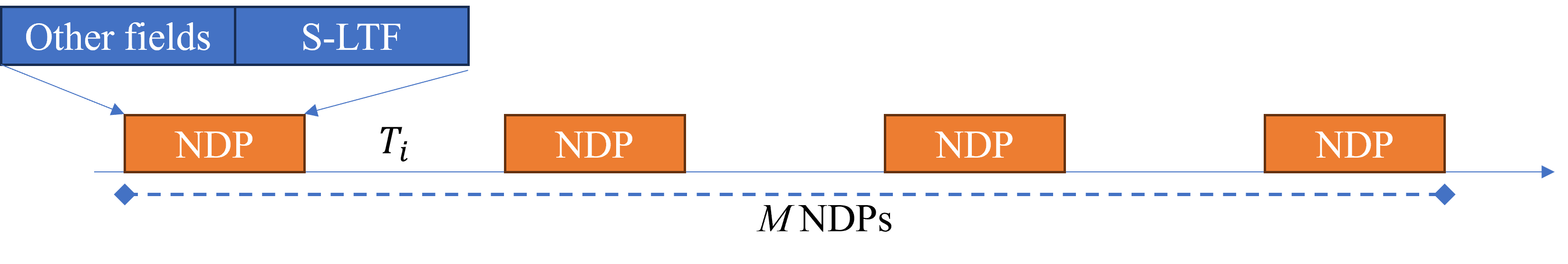}
    \caption{An overview of sensing measurement instance (SMI), where a null data packet (NDP) is composed of a sensing-long training field (S-LTF), which refers to the only OFDM symbol used for sensing, and other fields such as Legacy-LTF, are ignored for simplicity.}
    \label{fig:sensing_signal}
\end{figure}

In order to enable bistatic sensing, first the AP discovers the STAs equipped with 802.11ac/ax/be Wi-Fi chipsets during the sensing session setup (SSS). Then, the AP and the previously paired STAs fix their sensing parameters, such as the signal bandwidth, etc., during the sensing measurement setup (SMS). Finally, radar-like sensing takes place during the so-called sensing measurement instance (SMI). In this phase, STx and SRx exploit channel sounding protocols initially implemented to enable multi-user multi-input multi-output communication in Wi-Fi \cite{mumimo}. More specifically, STx transmits priorly known packets, called null data packets (NDPs), through the wireless channel, and the SRx estimates the radar channel transfer function (CTF) from the corresponding sensing-long training fields (S-LTFs) found in each NDP.\footnote{Depending on the amendment, S-LTFs have different names: VHT, HE, and EHT for 11ac, 11ax, and 11be, respectively. For simplicity, we refer to them as Sensing LTFs.} Hence, as illustrated in Fig.~\ref{fig:sensing_signal}, one can see the entire SMI as a pulsed OFDM-modulated radar scheme where each pulse corresponds to an NDP/S-LTF with PRI $T_i$.

Since the SSS and SMS phases described above take place over the air in bistatic geometry, a jammer can listen to these transmissions: i) if it is acting as a legitimate Wi-Fi device such as a neighboring AP, or an STA paired with the sensing AP\footnote{In Wi-Fi, the unicast and multicast management frames are protected so that eavesdropping and forging are avoided, hence, the need for acting as a legitimate device.}, or else ii) it can passively eavesdrop on the transmission of NDPs and estimate the WLAN Sensing parameters among a pre-determined set of parameters.\footnote{In any other non-standardized radar system, the radar parameters can take any value, which raises the need for DRFM. In WLAN Sensing, the size of the set of possible radar parameters is between 4 and 8 \cite{wlan_sensing2}.} Regardless, deceptive jamming becomes relatively straightforward.

\subsection{System Model}
\label{sec:ideal_system_model}

\begin{figure}
    \centering
    \includegraphics[width=\linewidth]{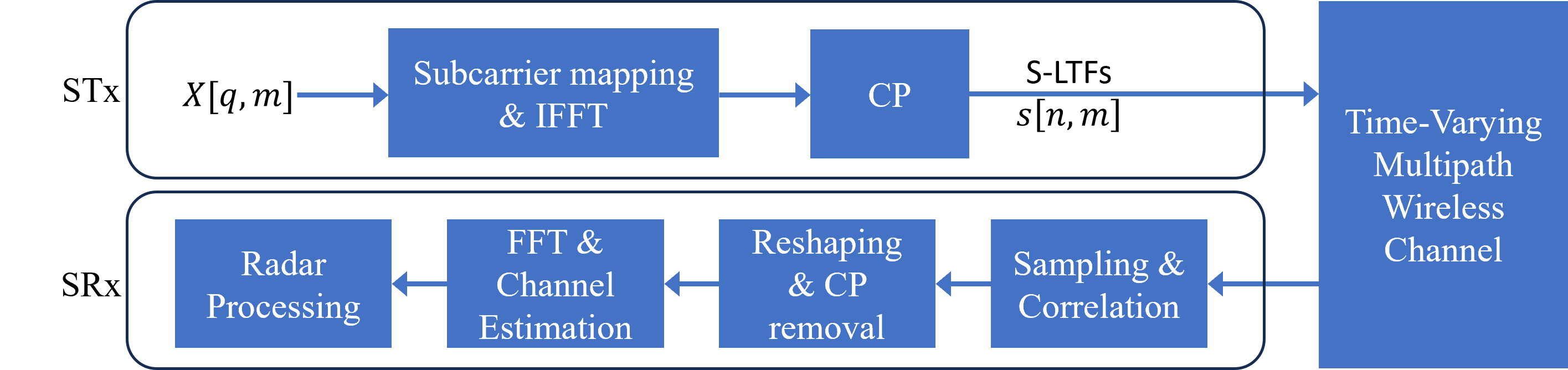}
    \caption{The block diagram for WLAN Sensing that shows the transmitter and the receiver processing stages up until the radar processing.}
    \label{fig:sensing_blocks}
\end{figure}

Since OFDM modulation is very well-known in the literature \cite{horlin},\cite{ofdm_basics3}, only a summary of the OFDM-based bistatic radar chain is provided. The OFDM modulation parameters are defined as follows: number of subcarriers $Q$,  number of samples in cyclic prefix (CP) $Q_{\text{cp}}$, system bandwidth $B$, sampling interval $T\!\!=\!\!1/B$, subcarrier spacing $\Delta_f\!\!=\!\!1/QT$, OFDM symbol duration $T_o\!\!=\!\!(Q+Q_\text{cp})T$ and PRI $T_i$ which is an integer multiple of $T_o$. In Fig.~\ref{fig:sensing_blocks}, a block diagram is provided where STx and SRx pursue the following stages: i) $X[q,m]$ contains standardized BPSK symbols on its subcarriers $q$ which are identical $\forall m$, and the inverse fast Fourier transform (IFFT) is computed over $X[q,m]$ along $q$ for the $m$-th S-LTF; ii) after adding of CP of length $Q_\text{cp}$, the S-LTFs are transmitted through the time-varying multipath channel; iii) SRx samples the received signal, finds the timing reference by using a correlator; iv) reshapes all the samples into parallel streams and removes the CP, and v) computes the fast Fourier transform (FFT) of each S-LTF symbol. 

Assuming that there is no carrier frequency offset between STx and SRx\footnote{Only residual CFO remains in S-LTF. The coarse CFO correction is already handled with Legacy-LTF which is received prior to S-LTF.}, and neither of the devices is mobile, the signal received on subcarrier $q$ of S-LTF $m$ is defined in the frequency domain as follows 
\begin{align}
    R[q,m] = H[q,m] X[q,m] + Z[q,m], \label{eq:true_recived_signal}
\end{align}
where $Z[q,m]$ is the additive white Gaussian noise (AWGN) at SRx, and $H[q,m]$ is the channel transfer function (CTF) to be estimated. The CTF is of the form 
\begin{align}
    H[q,m]=\sum_{p=0}^P \alpha_p e^{-j2\pi q\Delta_f(\tau_p-\tau_0)} e^{j2\pi mT_i f_p },\label{eq:channel}
\end{align}
where $P$ denotes the number of echoes, which are characterized by their amplitude $\alpha_p$, propagation delays $\tau_p$ with respect to the timing reference $\tau_0$, and Doppler frequency shifts $f_p$, while $p=0$ refers to the LOS between STx and SRx.

Assuming that the timing reference is STx line-of-sight (LOS), the estimated CTF is trivially written as 
\begin{align}
    \hat{H}[q, m] &= R[q,m]/X[q,m]
    \label{eq:ideal_ctf}
\end{align}
 from which  the RDM is obtained through a series of inverse discrete Fourier transforms (IDFT) over $q$ and discrete Fourier transforms over $m$ (DFT) as (for $l=0,\hdots,Q-1$ and $v=0,\hdots,M-1$)
\begin{align}
    & \hat{Y}[l,v] = \sum_{q=0}^{Q-1} \sum_{m=0}^{M-1} \hat{H}[q,m] e^{j2\pi \frac{q l}{Q}} e^{-j2\pi \frac{mv}{M}} \nonumber \\
    & = \sum_{p=0}^P \alpha_p \sum_{q=0}^{Q-1} e^{j2\pi \frac{q}{Q} (l-l_p)} \sum_{m=0}^{M-1} e^{-j2\pi \frac{m}{M} (v-v_p)}+Z[l,v]
    \nonumber \\ 
    & = \sum_{p=0}^{P} \alpha_p D_Q(l, l_p) D_M(v, v_p) + Z[l,v] \label{eq:ideal_rdm}
\end{align}
where $l_p = (\tau_p-\tau_0)/T$ and $v_p = T_i f_p$ correspond to the target propagation delay relative to the STx LOS and Doppler frequency shift, each normalized with respect to sampling interval and PRI, respectively. In general, neither $l_p $ nor $v_p$ are integers.\footnote{In an unlikely case where $l_p$ and $v_p$ are integers, the Dirichlet functions reduce to a Dirac delta function. However, $l_p$ and $v_p$ are rarely integers in reality, yielding sidelobes on the range and Doppler profiles which can be suppressed by windowing functions. } 
Moreover, $D_N(y, x)=e^{j\pi \frac{N-1}{N} (y-x)} { \sin(\pi(y-x)) }/{\sin(\pi {(y-x)}/{N})}$ corresponds to the Dirichlet kernel obtained by expanding the geometric series of Fourier transforms \cite{dirichlet}. 
Once an RDM is obtained, a constant false-alarm rate (CFAR) detector separates the target echo peaks from noise peaks \cite{principles}.



\section{Deceptive Jammer: Signal and Effects}
\label{sec:jammer_system_model}
In this section, we describe the deceptive jammer, and how it can be designed to deteriorate the WLAN Sensing performance on the target detection layer, i.e., at the output of the CFAR detector applied on an RDM, and on the tracking layer which is applied on successively obtained detection maps. We assume that the jammer can either listen to the sensing-related procedures between STx and SRx by acting as a legitimate device during SSS and SMS, or it can passively eavesdrop on the NDP/S-LTF transmissions to deduce the sensing parameters. In return, the jammer can tune its analog front-end for the specific sensing parameter, such as the carrier frequency, sampling rate, etc. We also assume that the jammer can transmit with more power than the STx. 

\subsection{Transmit Signal by the Jammer}
The artificial RDM generated by the jammer, which causes a single target at the SRx, is defined as follows:
\begin{align}
    & \Bar{Y}[l,v] = \Bar{\alpha} D_Q(l,\Bar{l}) D_M(v,\Bar{v})\label{eq:jammer_rdm}
\end{align}
with $\Bar{l} = \Bar{\tau}/T$ and $\Bar{v}=\Bar{f}T_i$. The bar symbol indicates the artificial parameters introduced by the jammer: $\Bar{\alpha}$, $\Bar{\tau}$ and $\Bar{f}$ correspond to the amplitude, propagation delay, and Doppler frequency shift of the phantom target, respectively. Here, $\Bar{\tau}>0, \Bar{\tau}\in\mathbb{R}$ to ensure that the phantom target has a positive range and $\Bar{f}\in\mathbb{R}$ can take any real value depending on the desired phantom target pattern to be forced at SRx.
Following the delay-frequency and Doppler-time dualities in \cite{durgin}, the CTF which yields the artificial RDM $\Bar{Y}[l,v]$ is obtained by computing the DFT over the range/delay axis and IDFT over speed/Doppler axis of \eqref{eq:jammer_rdm}, yielding
\begin{align}
    \Bar{H}[q, m] &= \Bar{\alpha} e^{-j2\pi q\Delta_f\Bar{\tau}} e^{j2\pi  mT_i \Bar{f}}.\label{eq:artificial_jammer_ctf}
\end{align}
After mapping each subcarrier with $\Bar{H}[q,m]$ and $X[q,m]$, the jammer signal takes the following form in the frequency domain
\begin{align}
    \Bar{S}[q,m] = \big(1+\Bar{H}[q,m]\big)X[q,m].
\end{align}
The first term in $(1+\Bar{H}[q,m])$ allows us to create an original copy of the S-LTF, i.e., untouched by the artificial RDM, to be forced as the timing reference at SRx. The second term allows us to pre-modulate the OFDM spectrum. Finally, computing the IDFT over $q$ yields
\begin{align}
    \Bar{s}[n,m] = \sum_{q=0}^{Q-1} \Bar{S}[q,m] e^{j2\pi \frac{qn}{Q}}, n = 0,\hdots,Q-1.
\end{align}
The first and second dimensions of $\Bar{s}[n,m]$ correspond to pre-modulated OFDM symbols and linearly increasing phase shifts for the Doppler profiles, respectively. After adding the CP to each OFDM symbol, the jammer transmits $\Bar{s}[n,m]$ in the time domain.

\subsection{Signal Received at SRx}
The channel between the jammer 
and the SRx is
\begin{align}
    H'[q, m] &= \sum_{\rho=0}^{P'} \alpha'_{\rho} e^{-j2\pi q\Delta_f\tau'_{\rho}} e^{j2\pi mT_i f'_\rho },
    \label{eq:jammer_ctf}
\end{align}
where $P'$, $\alpha'_\rho$, $\tau'_\rho$, and $f'_\rho$ correspond to the number of echoes, attenuation, propagation delay, and Doppler frequency shift, respectively, while $\rho=0$ refers to the LOS between the jammer and SRx. 

The jamming signal perceived by SRx thus takes the following form in the frequency domain
\begin{align}
    R'[q,m] &= H'[q,m] S[q,m] = H'[q,m] (1+\Bar{H}[q,m]) X[q,m] \nonumber \\
    &= \sum_{\rho=0}^{P'} \alpha'_{\rho} e^{-j2\pi q\Delta_f \tau'_{\rho}} e^{j2\pi mT_i f'_\rho} X[q,m] \label{eq:jammer_received_signal} \\
    &+\sum_{\rho=0}^{P'} \alpha'_{\rho}\Bar{\alpha} e^{-j2\pi q\Delta_f (\tau'_{\rho}+\Bar{\tau})} e^{j2\pi mT_i (f'_\rho+\Bar{f}) }X[q,m]. \notag
\end{align}
Since the jammer transmits an original copy of the S-LTF, signals from the true targets between the jammer and SRx are also received, modeled by the first sum in \eqref{eq:jammer_received_signal}. Note that, since the position of the jammer is generally different than the STx, the radar parameters of the targets in \eqref{eq:jammer_received_signal} are different than those in \eqref{eq:ideal_ctf}, i.e., $\alpha_p\neq\alpha'_p$, $\tau_p\neq\tau'_p$ and $f_p\neq f'_p, \forall p$. On the other hand, the multipath channel between SRx and the jammer generates multiple copies of the phantom target, modeled by the second sum in \eqref{eq:jammer_received_signal}. 

Putting the legitimate signal from the STx and the signal from the deceptive jammer together, the SRx observes
\begin{align}
    & R_{o}[q,m] = R[q,m] + R'[q,m] \nonumber \\
    &= \underbrace{H[q,m] X[q,m]}_{\text{sensing of STx-SRx channel}} 
    + \underbrace{H'[q,m]X[q,m]}_{\text{sensing of jammer-SRx channel}} \nonumber \\
    &+ \underbrace{H'[q,m]\Bar{H}[q,m] X[q,m]}_{\text{artifical RDM between jammer and SRx}} +Z[q,m].
\end{align}

\subsection{Different Cases for Jammer Signal Time-of-Arrival}
\label{sec:jammer-cases}
\begin{figure}
    \centering
    \includegraphics[width=\linewidth]{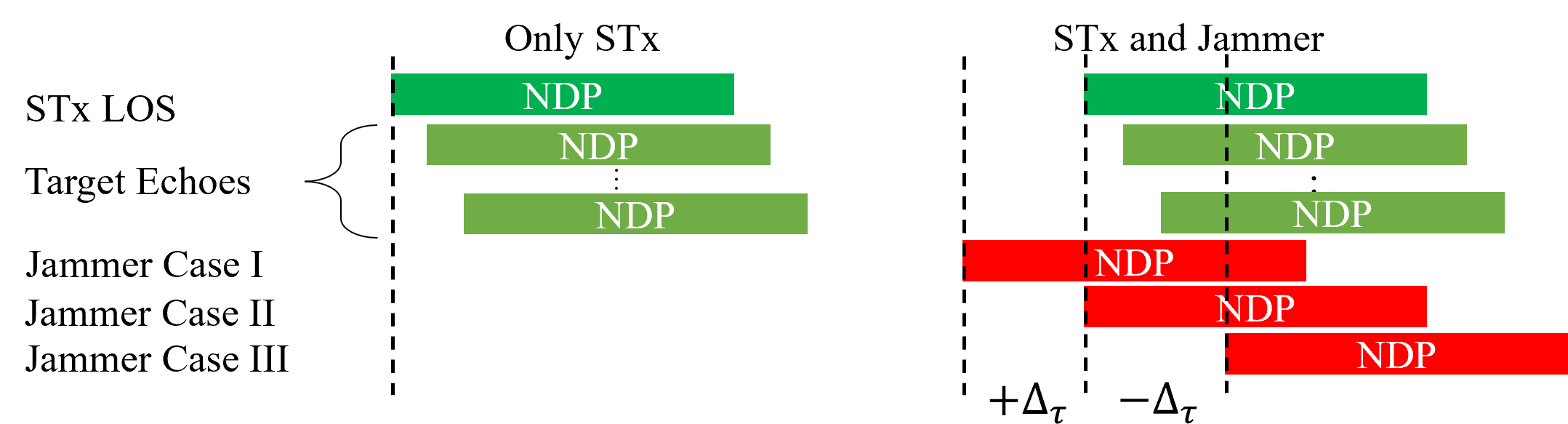}
    \caption{An illustration of the echoes when only the STx is active vs. both the STx and the jammer are active.}
    \label{fig:ideal_vs_jammed}
\end{figure}

The time difference of arrival between the true sensing signal and the jamming signal has great consequences at SRx as illustrated in Fig.~\ref{fig:ideal_vs_jammed}. Ideally, the timing reference, obtained after the correlation at SRx, will be the S-LTF that propagates through STx LOS since it exhibits the largest amplitude at SRx. However, the timing reference can potentially be triggered by another S-LTF transmitted by the jammer. The time difference of arrival between the true and jamming signals is defined as $\Delta_\tau=\tau'_0-\tau_0\pm\Bar{\epsilon}$. Here, $\Bar{\epsilon}$ can be a deterministic variable that is used by the jammer to time-align its signals with the true sensing signals depending on the desired effects. However, the jammer needs to know $\tau'_0$ and $\tau_0$ to have full control over $\Delta_\tau$. On the other hand, $\Bar{\epsilon}$ can be a random scenario-specific variable if neither $\tau'_0$ nor $\tau_0$ are known. In this case, the type of jamming effects perceived by the SRx cannot be guaranteed. Assuming that the jamming LOS is stronger than STx LOS, the different consequences depending on the value of $\Delta_\tau$ are summarized as follows: 
\begin{itemize}
    \item \emph{Jammer Case I:} The jamming LOS signal arrives earlier than the STx LOS. The true RDM will be positively shifted based on $\Delta_\tau / T$, i.e., the true echoes will appear at further distances. The phantom target, and its multipath components, will appear at $\Bar{\tau} + \tau'_\rho - \tau’_0,\;\rho=1,\hdots, P'$.
    \item \emph{Jammer Case II:} In an unlikely scenario where the STx LOS and the jamming signal arrive simultaneously at the SRx, both the true and phantom targets will be present near the zero range, with both of their LOS paths appearing exactly at the zero range.
    \item \emph{Jammer Case III:} The jamming signal arrives later than the STx LOS. The true RDM will be negatively shifted based on $\Delta_\tau / T$, i.e., the true echoes will appear to be at closer distances. This can potentially destroy the subcarrier orthogonality on the true sensing signal if it is sampled beyond its CP. The phantom target, and its multipath components, will appear as in Jammer Case I.
\end{itemize}
The combined RDM perceived by the SRx can be written as follows
\begin{align}
    \hat{Y}_o[l,v] &= \sum_{p=0}^P \alpha_p D_Q(l,l_p\pm\Delta_\tau / T) D_M(v,v_p)  + Z[l,v] \nonumber \\
    &\hspace*{-1cm}+\sum_{\rho=0}^{P'} \alpha'_\rho D_Q(l,l'_\rho) D_M(v,v'_\rho) + \sum_{\rho=0}^{P'} \alpha'_\rho \Bar{\alpha} D_Q(l,\Bar{l}_\rho) D_M(v,\Bar{v}_\rho) \label{eq:jammed_rdm}.
\end{align}
The first sum in \eqref{eq:jammed_rdm} models the true RDM between STx and SRx, range shifted by $\Delta_\tau/T$ number of range gates due to the forced timing reference. The second sum corresponds to the RDM between the jammer and SRx, also with the true target. However, its normalized range $l'_\rho=\tau'_\rho/T$ and Doppler frequency $v'_\rho=T_if'_\rho$ differs from the true RDM since the location of the jammer is different than STx, as shown in Fig.~\ref{fig:scenario_geometry}. Finally, the third sum corresponds to the artificial RDM propagated through the multipath channel, with $P'$ number of phantom targets at the normalized ranges $\Bar{l}_\rho = (\tau'_\rho+\Bar{\tau})/T$ and Doppler frequencies $\Bar{v}_\rho = (v'_\rho + \Bar{v})T_i$.

The model provided in \eqref{eq:jammed_rdm} corresponds to a single snapshot, yielding a single RDM. However, modern radar systems use multiple snapshots to track the targets over time. Thanks to the flexibility of the proposed jamming method, RDMs with a target (or potentially multiple targets) that is moving according to Newtonian kinematics can digitally be designed and transmitted by the jamming signal over multiple snapshots. 

\section{Performance Evaluation}
\label{sec:numerical_analyses}
In this section, numerical analyses illustrate the vulnerability of WLAN Sensing against deceptive jamming, and experimental results are provided for validation. 

\subsection{Scenario and Parameters}
Jammer Cases I and III from Section \ref{sec:jammer-cases} have been implemented. To ensure that all the effects described in Sections \ref{sec:ideal_system_model} and \ref{sec:jammer_system_model} are taken into account, the entire simulation chain shown in Fig.~\ref{fig:sensing_blocks} is simulated, and the simulation parameters are summarized as follows: $B=80$ MHz, $Q=1024$, $Q_\text{cp}=64$, $M=128$, $T_i=2\text{ms}$, and the STx transmit power is 23 dBm.

\begin{figure}
    \centering
\includegraphics[width=\linewidth]{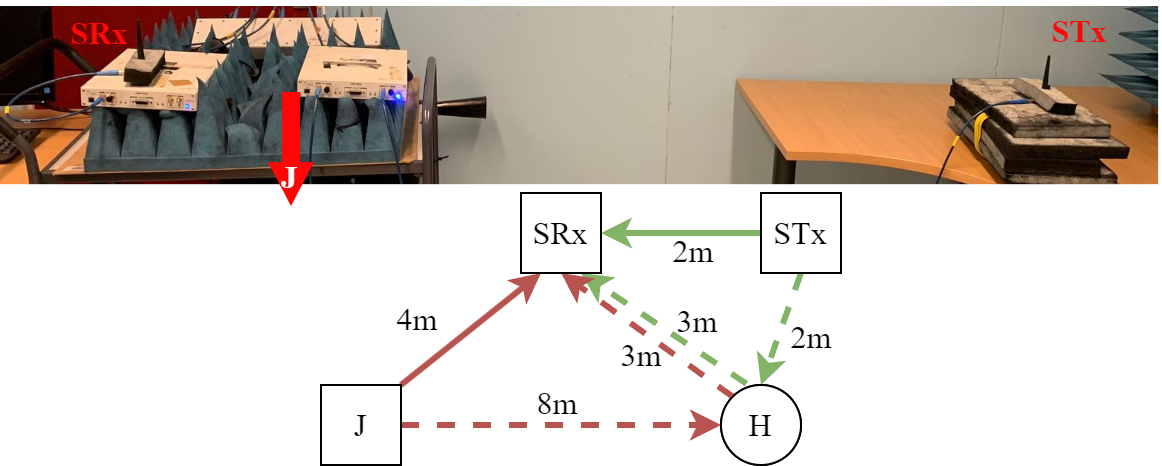}
    \caption{The picture and the diagram of the experimental setup. The jammer (J) antenna is behind the camera. Solid and dashed lines correspond to the LOS and reflections from the target (H), respectively.}
    \label{fig:exp_scenario}
\end{figure}

For the experimental results, a real-life scenario with a moving person in a room is considered, as shown in Fig.~\ref{fig:exp_scenario}. Two USRP X310s are used to emulate the devices: one for the STx and the Jammer, another for the SRx, while we maintain the parameters used in our simulations, with 30 dBm transmit power at the jammer.

\subsection{Simulation Results}
\begin{figure}
    \centering
    \includegraphics[width=\linewidth]{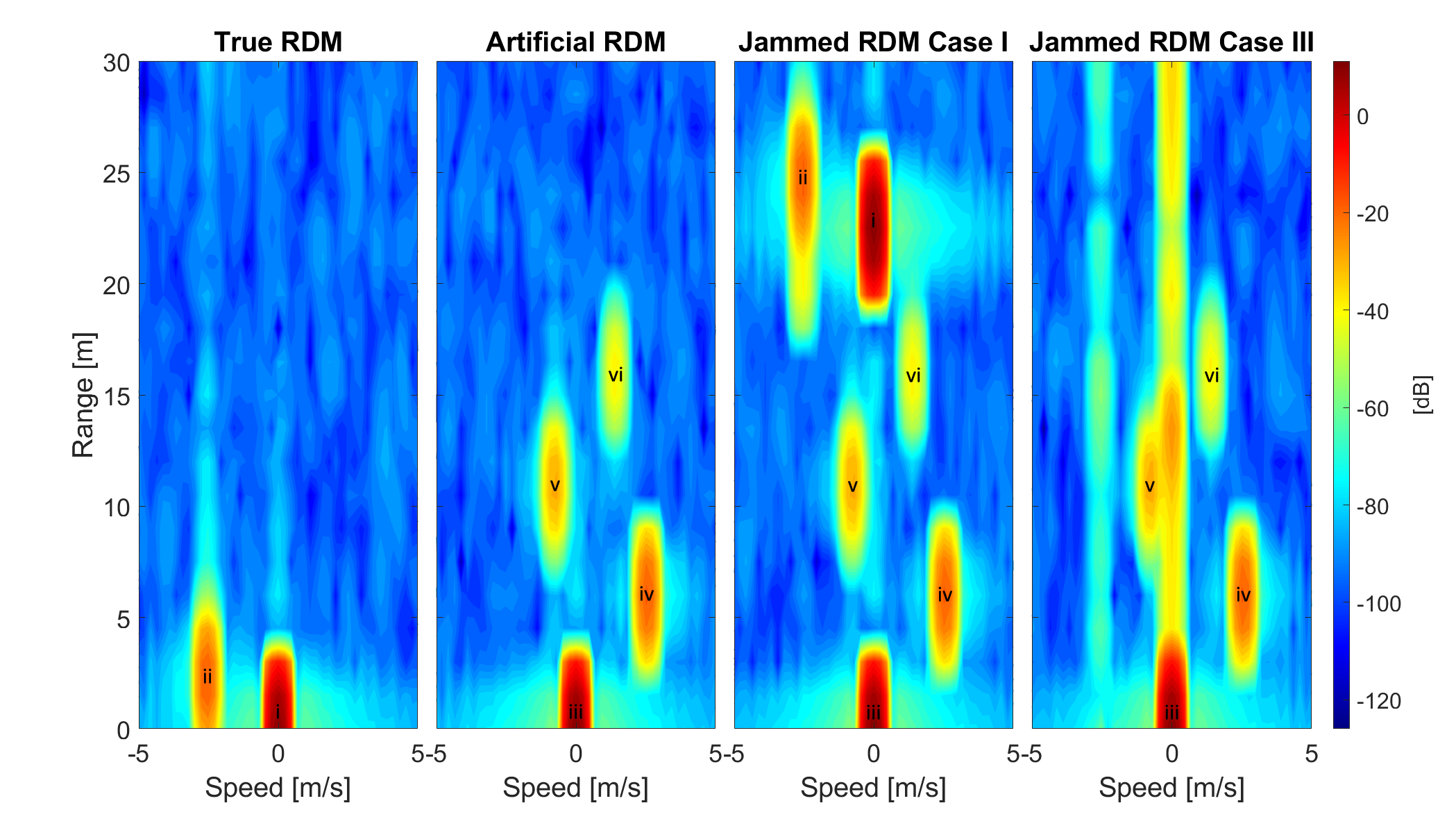}
    \caption{Simulation: Jammer Cases I and III when the timing reference is triggered by the jammer LOS.}
    \label{fig:stronger_jammer}
\end{figure}

In Fig.~\ref{fig:stronger_jammer}, four RDMs are provided, which we now discuss in detail:
\begin{itemize}
    \item \emph{True RDM:} The first RDM is obtained when only the true sensing echoes are received by SRx. Peak (i) is the timing reference, while peak (ii) is the true mobile target. 
    \item \emph{Artificial RDM:} The second RDM is obtained when only the jamming signal is received by SRx. Peak (iii) corresponds to the jammer timing reference, while peaks (iv), (v), and (vi) correspond to the phantom target, the true target, and the phantom target signal reflected from the true target, respectively. Notice that the true target has different characteristics compared to (ii) since the jammer is positioned differently than the STx, as depicted in Fig.~\ref{fig:scenario_geometry}. 
    \item \emph{Jammed RDM Case I:} The third RDM is obtained when SRx first receives the jamming signal and then the true sensing signal, i.e., Jammer Case I with $\Delta_\tau / T= 16$. As pointed out in \eqref{eq:jammed_rdm}, SRx perceives a linear combination of the artificial RDM and the true RDM range shifted by about 30 meters. In case $\Delta_\tau/T$ is much larger, the true target will appear even further away and potentially will be ignored by the tracking layer due to its distance.
    \item \emph{Jammed RDM Case III:} The last RDM corresponds to the Jammer Case III with $\Delta_\tau / T = -70$. In this case, the RDM exhibits the phantom target, the timing reference peak from the jammer, and ridges along the range dimension. These ridges are due to the loss of subcarrier orthogonality on the true sensing symbols since they are sampled beyond their CP. Moreover, the fact that the true target is not visible at all introduces an even bigger problem than the previous one since there is nothing to detect/track at all.

\end{itemize}

\begin{figure}
    \centering
    \includegraphics[width=\linewidth]{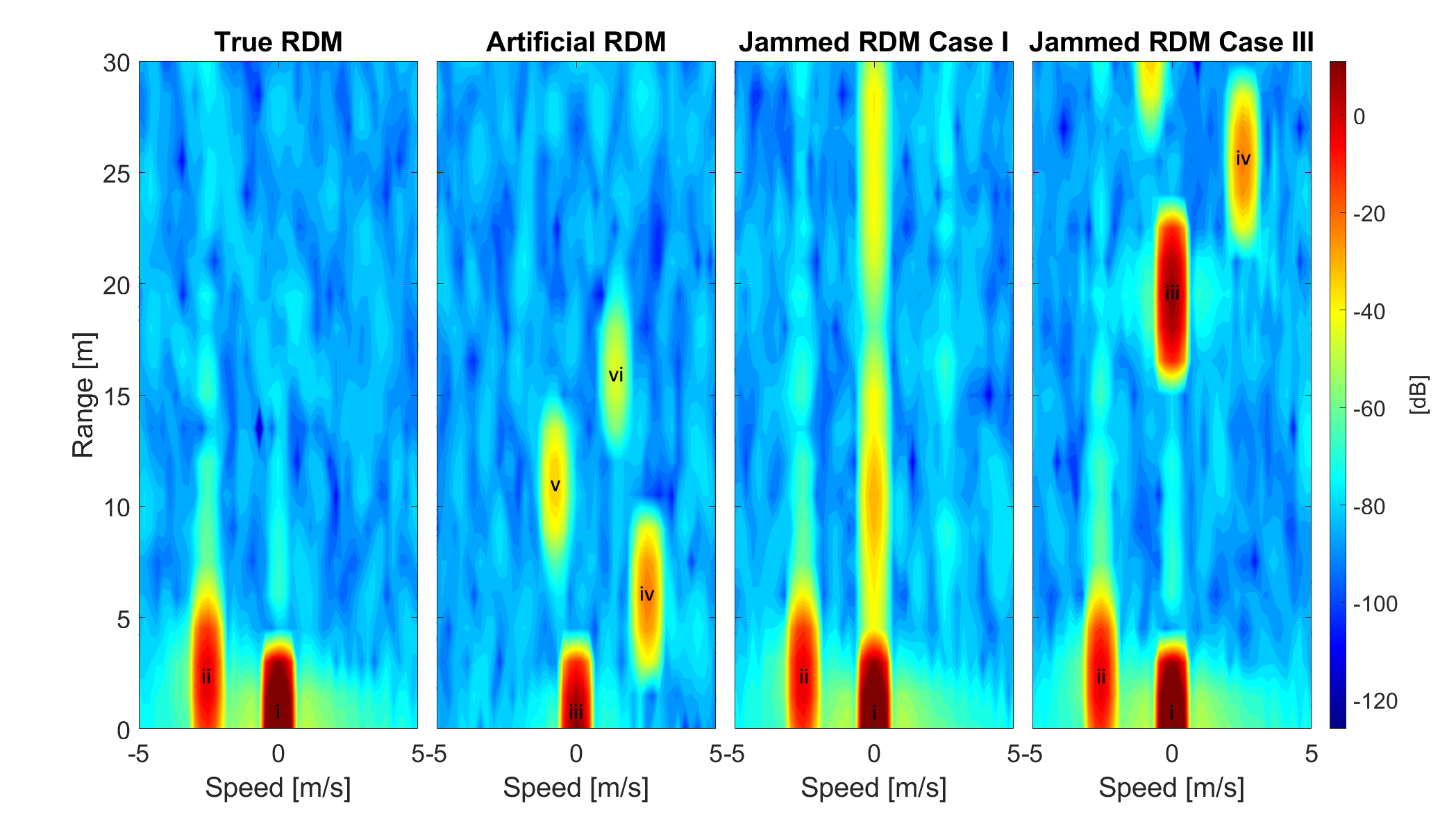}
    \caption{Simulation: Jammer Cases I and III when the timing reference is triggered by the STx LOS.}
    \label{fig:weaker_jammer}
\end{figure}

For the sake of completeness, Fig.~\ref{fig:weaker_jammer} is provided where the amplitude of STx LOS is larger than the jammer LOS, hence, the timing reference is triggered by the STx LOS. In this case, the previously mentioned effects are reversed. When the jamming signal arrives earlier than STx LOS, the phantom target that it carries is not visible on the jammed RDM. Instead, the jammer symbols yield the ridges due to the lost orthogonality among its subcarriers. On the other hand, when the jamming signal arrives later than STx LOS, the phantom target as well as the jammer LOS are visible at further distances than intended.

\subsection{Experimental Results}
\label{sec:exp_res}

\begin{figure}
    \centering
\includegraphics[width=\linewidth]{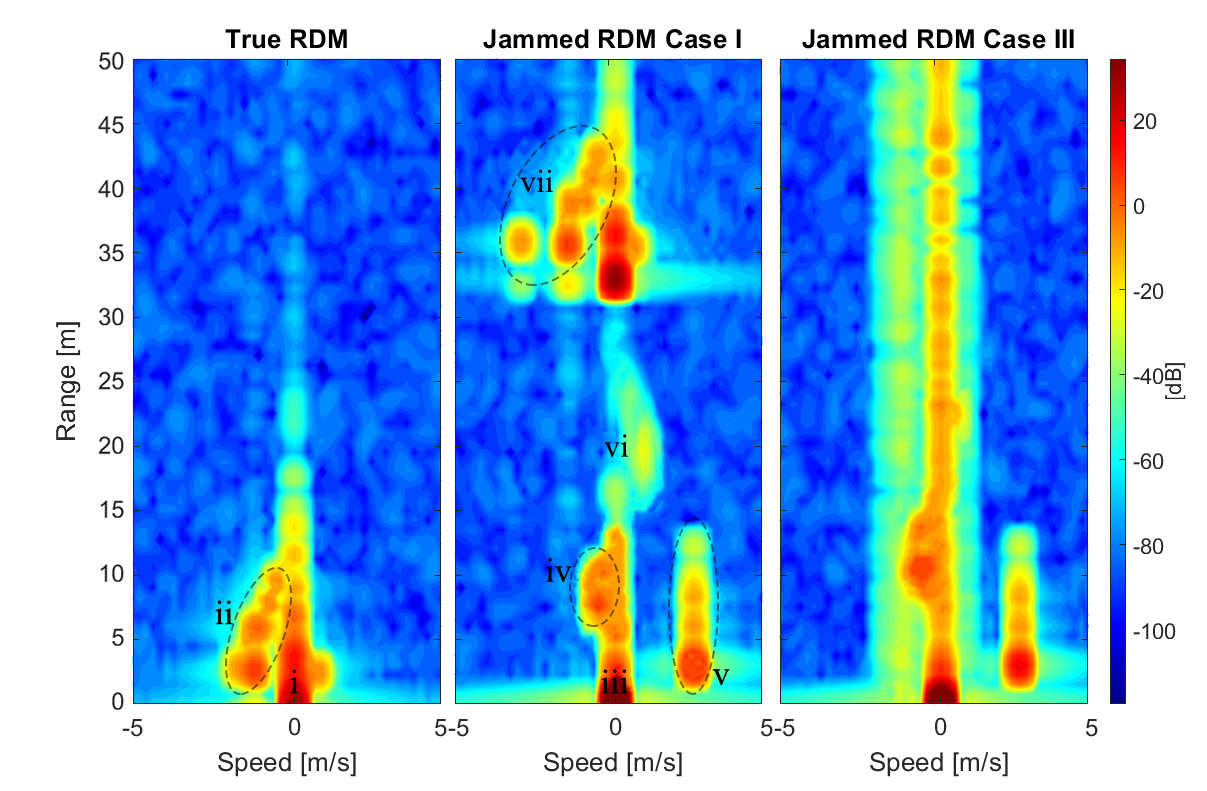}
    \caption{Experiment: Indoor results with a human target considering Jammer Cases I and III.}
    \label{fig:exp_res}
\end{figure}

In Fig.~\ref{fig:exp_res}, three RDMs are provided, which we again discuss in detail: 
\begin{itemize}
    \item \emph{True RDM:} The true RDM is obtained when only the STx is active in the environment. As opposed to the simulation results from Fig.~\ref{fig:stronger_jammer}, the RDMs possess a few differences. First, the static clutter (i) is visible at 0 m/s, extending until 40 meters range. Second, the true mobile target (ii) is also subject to multipath, therefore, there are ghost targets\footnote{A phantom target refers to a target digitally generated by the jammer, whereas a ghost target refers to the multipath components of true or phantom targets.} until 20 meters. However, since the ghost targets do not move in the same direction as the true target, they yield slightly different Doppler frequency shifts. 
    \item \emph{Jammed RDM Case I:} 
    The second RDM corresponds to the Jamming Case I, where both the STx and the jammer are active. The jammed RDM exhibits the static clutter (iii), the true target (iv), the phantom target (v), and the phantom-to-true target (vi). As predicted by the numerical results, the artificial RDM is the one that is present at short ranges. Meanwhile, the true RDM (vii) (which has slightly different Doppler characteristics compared to the first RDM since it is a new realization) is range-shifted. The most important difference is how the phantom target (v) reacts to the multipath conditions. As opposed to the true target, the artificially generated target is not an actual object moving in the environment, hence, it does not experience the same physical effects. Instead, its ghost targets appear at the same speed gate at further distances. 
    \item \emph{Jammed RDM Case III:} Finally, the third RDM corresponds to the Jamming Case III with a new experimental realization. In this case, two ridges along the range are visible, one for the static clutter and the other for the true mobile target. As pointed out earlier, when the true sensing symbols are sampled far beyond their CP, the orthogonality on their subcarriers is completely lost, yielding such sidelobes along the range profile \cite{ofdm_basics3}. Apart from the true target peaks, the sensing system is almost completely deceived since the phantom target is the only mobile peak that appears on the RDM.
\end{itemize}
Overall, there is a good match between simulation and experimental results, indicating that the models from Section \ref{sec:jammer_system_model} are valid. 

\section{Conclusion}
\label{sec:conclusion}
In this paper, an important but overlooked fact regarding the OFDM-based JCAS systems and their proneness against deceptive jamming was studied. Since JCAS is already taking shape within WLAN Sensing, we specifically focused our analyses on the parameters, the protocols, and the waveforms used in WLAN. We have shown that OFDM makes it very easy to digitally generate realistic RDMs for deceptive jamming, and the underlying methods behind bistatic radar processing further help the deceptive jammer to either push the true targets away along the range dimension or completely eliminate them from the RDM. Our experiments demonstrated that the deceptive jammer is easy to implement, and its consequences should raise concerns regarding the safety and security applications foreseen in WLAN Sensing, and more generally, in OFDM-based JCAS. We conclude that to guarantee robust and future-proof JCAS systems, electronic-counter-countermeasure techniques must be studied, developed, and included in JCAS standardization. Two important research questions remain to be answered: how deceptive jamming can be applied in a 6G context and how 6G can be designed to be less susceptible than WLAN. 


%
\end{document}